\documentclass[aps,preprint,prx,a4paper]{revtex4-1}
\usepackage{CJK}
\usepackage{amsmath,amssymb,graphicx,siunitx}
\usepackage{chemformula}
\usepackage[utf8]{inputenc}

\newcommand{\pdiff}[3][]{\frac{\partial^{#1} #2}{\partial #3^{#1}}}
\newcommand{\We}{\mathrm{We}}
\newcommand{\Rey}{\mathrm{Re}}
\newcommand{\St}{\mathrm{St}}

\begin{document}
\begin{CJK*}{UTF8}{}
\title{Cold granular targets slow the bulk freezing of an impacting droplet}

\author{Song-Chuan Zhao \CJKfamily{gbsn}(赵松川)}
\email[]{songchuan.zhaon@outlook.com}
\author{Hao-Jie Zhang \CJKfamily{gbsn}(张浩杰)}
\author{Yudong Li \CJKfamily{gbsn}(李宇栋)}

\affiliation{State Key Laboratory for Strength and Vibration of Mechanical Structures, School of Aerospace Engineering, Xi'an Jiaotong University, Xi'an 710049, China}

\begin{abstract}
	When making contact with an undercooled target, a drop freezes. The colder the target is, the more rapid the freezing is supposed to be. In this research, we explore the impact of droplets on cold granular material. As the undercooling degree increases, the bulk freezing of the droplet is delayed by at least an order of magnitude. The postponement of the overall solidification is accompanied by substantial changes in dynamics, including the spreading-retraction process, satellite drop generation, and cratering in the target. The solidification of the wetted pores in the granular target primarily causes these effects. Owing to the small size of pores, solidification there is sufficiently fast to match the characteristic timescales of the impact dynamics at moderate undercooling degrees. As a result, the hydrophilic impact appears `hydrophobic', and the dimension of the solidified droplet shrinks. A monolayer of cold grains on a surface can reproduce these consequences. Our research presents a potential approach to regulate solidified morphology for subfreezing drop impacts. It additionally sheds light on the impact scenario of strong coupling between the dynamics and solidification.
\end{abstract}

\maketitle
\end{CJK*}

\section{Introduction} 
Drop impact on subfreezing targets causes ice accretion. Depending on the perspectives, it could cause concern in freezing rain~\cite{Jones1998,Laforte1998} and aircraft icing~\cite{Dalili2009,Baumert2018} or benefit others, such as spraying and additive manufacturing techniques~\cite{Gao1997,Visser2015}. In recent years, this topic has gained significant attention and interest~\cite{Schutzius2015}. Regulating the solidification morphology is the general goal of the research, either inhibiting ice formation~\cite{Mishchenko2010,Maitra2014} or manipulating the adhesion~\cite{Golovin2016}. Due to the complexity of the context, a general framework of the solution is still an open question~\cite{Schutzius2015,Kreder2016}. A comprehensive understanding of the impact process under various conditions is thus necessary to develop effective strategies for controlling the icing morphology.

The dynamics of isothermal droplet impact are governed by the interplay of forces resulting from the inertia of the impacting droplet, surface tension and viscosity. Subfreezing impacts, where the temperature of the droplet ($T_d$) or of the target ($T_g$) or both are lower than the melting point of the impacting liquid $T_m$, induce solidification in addition. Thermal diffusion governs solidification, which is typically slower than impact dynamics. For impact on a solid substrate of extreme undercooling degree, \textit{e.g.}, $\Delta T = T_m - T_g\sim 100\si{K}$, the thermal diffusivity of water barely matches its viscosity~\cite{Thievenaz2019,Thievenaz2020,Gielen2020,Schremb2018}. Therefore, the impact dynamics and the freezing process are largely decoupled. For instance, droplet spreading does not show apparent dependence on the undercooling degree~\cite{Kant2020,Kant2020a}, even though early-stage nucleation may introduce subtle dynamic features~\cite{DeRuiter2018,Kant2020}. The role of solidification processes on the impingement morphology is only considered seconds after the impact~\cite{Shang2019,Hu2020,Fang2021}, where bulk solidification is realized, and thermal stresses lead to cracking~\cite{Ghabache2016,Song2020} or bending~\cite{DeRuiter2018,Fang2022}. 
Regulating the morphology is thus challenging due to the weak coupling between the solidification and overall dynamics.

In this research, we study a scenario where the interplay between solidification and the impact dynamics occurs at the moderate undercooling degree (\textit{e.g.,} $\Delta T\sim 30\si{K}$ for water), droplet impact on subfreezing granular layers. Compared with conventional impacts, a granular target introduces new features. The liquid-grain mixing is crucial for the dynamics and results in a wide variety of impingement morphologies~\cite{Katsuragi2010,Delon2011,Emady2013}. A recent work~\cite{Lin2022} reports that this mixing process significantly enhances the vaporization of a droplet impacting hot granular material. We will show that such enhancement of the phase transition is present locally here as well; however, it leads to the opposite effect in bulk, \textit{i.e.,} delaying the overall freezing of the droplet. This delay is accompanied by a pronounced decrease in the dimension of the frozen residual and, thus, its contact area with the target. In essence, the impact on hydrophilic grains appears \textit{hydrophobic} at lower temperatures. Moreover, the frozen layer shields shearing erosion on the substrate during impact.

\section{Experimental protocol} 
Water drops with a diameter $D_{0} = 3.30 \pm 0.03 \si{mm}$ are generated through a nozzle by a syringe pump. The releasing height varies to achieve a range of the impact velocity $U=1.7\si{m/s} - 3.2\si{m/s}$. The droplets are dyed for visualization and maintained at a temperature of $T_{d} = 0\si{\celsius}$ before impact. The target is a granular packing of hydrophilic ceramic spherical particles (Yttrium stabilized Zirconium Oxide, purity $ \geq 94.6 \% $) with a diameter $d_g=163.6\si{\micro\meter}$. The packing density, $\phi$, defined by particle volume fraction, is set at $\phi=0.59$. Results remain qualitatively consistent for densities within the range $\phi=\numrange[range-phrase = \text{--}]{0.58}{0.61}$. The granular packing is initially fluidized with cold nitrogen gas, lowering the grain temperature, $T_g$, to approximately $-70\si{\celsius}$. After the nitrogen flow ceases, $T_g$ increases gradually. Impact experiments are then conducted at the desired $T_g$. The undercooling degree $\Delta T = T_m - T_g$ is defined with $T_m = 0\si{\celsius}$ as the melting temperature of the liquid. Constants, such as the density of the liquid and grains, $\rho_l$ and $\rho_{g}$, and the surface tension coefficient, $\gamma$, are detailed in Methods.

\begin{figure}
	\centering
	\includegraphics{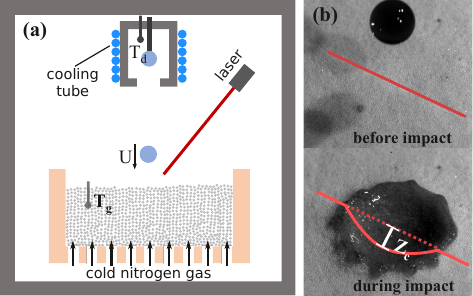}
	\caption{\label{f.exp} (a) Schematic of the experimental setup. (b) An illustration of laser profilometry, measuring the deformation of the granular packing. Solid lines indicate the laser line.}
\end{figure}

The setup is placed inside a sealed chamber filled with dry nitrogen gas to avoid frost formation, as depicted schematically in Fig.~\ref{f.exp}a. Each impact is captured by a high-speed camera (Photron SA-Z), while a secondary camera records the substrate deformation via laser profilometry, as illustrated in Fig.~\ref{f.exp}b. This method, previously validated~\cite{Zhao2015,Zhao2017,Zhao2019,deJong2017,deJong2021}, primarily focuses on the maximum crater depth, $Z_c$, during the impact. Readers may refer to Methods for details of the experimental protocol and measurement precision.

\section{Results}
\subsection{Phenomena} 

\begin{figure*}
	\centering
	\includegraphics{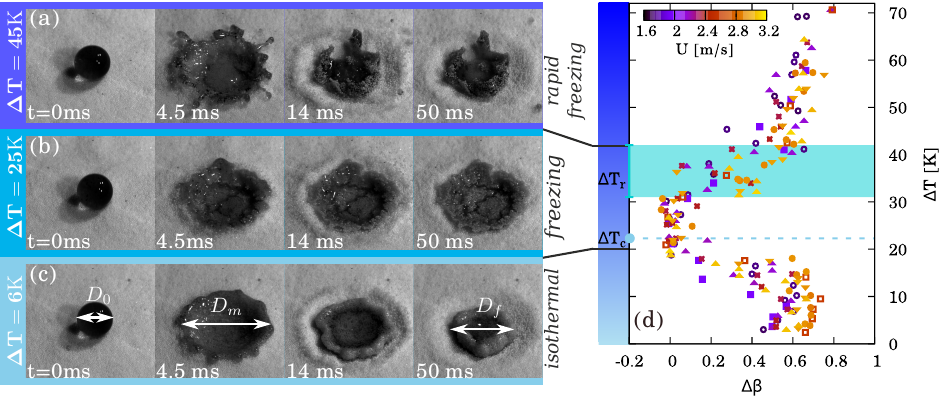}
	\caption{\label{f.pheno} Impact events of $U=2.15\si{m/s}$ and three undercooling degrees (a-c), exemplifying typical behavior in the \textit{isothermal}, \textit{freezing} and \textit{rapid-freezing} undercooling regimes. Multimedia views are available in the supplementary material. (d) For various $U$, the retraction degree, $\Delta\beta=(D_m-D_f)/(D_m-D_0)$, develops nonmonotonically with the undercooling degree $\Delta T$. The dot (the bar between $32\si{K}$ and $40\si{K}$) on the axis of $\Delta T$ indicate the critical undercooling degree $\Delta T_c$ ($\Delta T_r$) in theory.}
\end{figure*}

Figure~\ref{f.pheno}a-c illustrate typical impact events for $U=2.15\si{m/s}$. The undercooling degree of the granular target, $\Delta T$, significantly alters the impact and retraction process of the droplet. For small $\Delta T$, the impact dynamics essentially replicate the isothermal impact at room temperature~\cite{Zhao2015,Zhao2019}, where the droplet spreads, mixes with grains and retracts after reaching the maximum spreading diameter, $D_{m}$. Any unmixed liquid sinks into the porous substrate afterward. In contrast, for $\Delta T\approx 20\si{K}$, the retraction is completely suppressed, and the mixture consists of icy patches connected by liquid films (Fig.~\ref{f.pheno}b). The remaining liquid solidifies after the dynamics cease, with little penetrating the substrate. For $\Delta T>30\si{K}$, an icy patch forms at the impact center, similar to moderate $\Delta T$, but fewer grains are mixed with the spreading droplet. Instabilities of the rim of the liquid film develop into fingering and drop fragments. It is remarkable that the consequent retraction re-occurs. After the retraction, approximately half of the droplet volume remains liquid, and the solidification of the droplet is delayed till 40-50 seconds, at least one order of magnitude longer compared to the intermediate subfreezing impact in Fig.~\ref{f.pheno}b and impacts on a subfreezing surface~\cite{Shang2019,Hu2020}. Namely, a colder substrate counterintuitively prolongs the overall freezing time. As only $\Delta T$ varies, the observation likely arises from the interplay between solidification and the impact dynamics.

A retraction degree, $\Delta\beta=(D_m-D_f)/(D_m-D_0)$, is introduced to characterize the overall spreading-retraction dynamics, where $D_f$ is the diameter of the residual at $t=50\si{ms}$ (Fig.~\ref{f.pheno}c)~\footnote{The shape of residual is not necessarily symmetric in azimuth. The transversal dimension through the impact center is always used here.}. In Fig.~\ref{f.pheno}d we plot $\Delta\beta$ with $\Delta T$ for various $U$. The variation trend in $\Delta \beta$ remains consistent across the range of $U$ explored. There are three regimes of $\Delta\beta$: two plateaus at the low and high undercooling degrees and a basin of $\Delta\beta\approx 0$ in-between, corresponding to the three typical impact processes shown in Fig.~\ref{f.pheno}a-c. The non-monotonic evolution of $\Delta\beta$ confirms that the retraction of the droplet is suppressed first by increasing $\Delta T$ and then returns at higher undercooling degrees for the range of $U$ explored. Henceforth, we refer to the three regimes as the \textit{isothermal}, \textit{freezing}, and \textit{rapid-freezing} regimes. 

Despite the distinct dynamics observed across the three undercooling regimes, a common underlying mechanism is shared among them. Prior research on drop impact on granular targets~\cite{Zhao2015,Zhao2017,Zhao2019} has highlighted the critical role of liquid-grain mixing in shaping dynamics, such as droplet spreading, retraction, splashing and the final morphology of the mixture. Building on this understanding, we suggest that at higher undercooling degrees, the solidification of the liquid penetrating the pores of the granular substrate hinders further droplet-grain mixing, which in turn substantially alters the dynamics and the final morphology of the solidified droplet. Indeed, as the undercooling degree is enhanced, the dramatically varied impact behavior is accompanied by an increasing amount of pure liquid remaining after impact, thus less mixing (see Fig.~\ref{f.pheno}a-c and Movie S1-S3). The suppression of mixing is also consistent with the observation of freezing delay. Since mixing creates contact areas between the droplet and the heat sinks, cold grains, a sufficient reduction of mixing could offset the increase of the undercooling degree of the substrate. Therefore, the freezing of the entire droplet is instead postponed. 

\begin{figure}
	\centering
	\includegraphics[width=8.4cm]{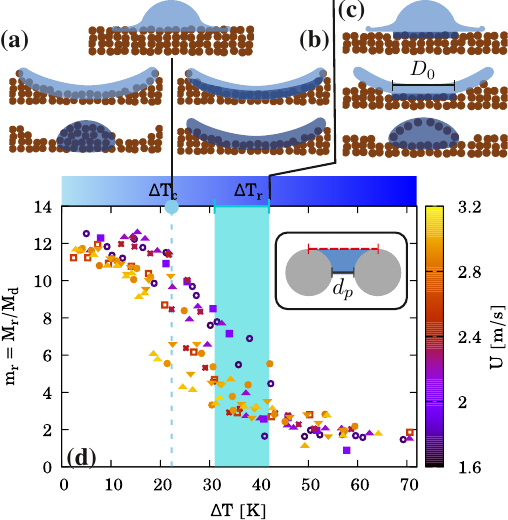}
	\caption{\label{f.res_mass} (a-c) Schematics of impact dynamics in three undercooling regimes. From top to bottom: the early impact, the maximum extension state, and the final morphology. Darker blue areas indicate the frozen portion. (d) The mass ratio between the residual and the droplet versus $\Delta T$ for various impact speeds $U$. $\Delta T_c$ and $\Delta T_r$ denote the theoretically calculated critical undercooling degrees. Inset: A schematic showing the dependence of the effective pore dimension on liquid penetration depth. The cross-section dimension on the packing surface (red dashed line) is larger than the average distance between grains, $d_p$, within the packing.
	}
\end{figure}

Quantitative evidence is given by the residual mass, $M_r$, measured two minutes after the impact. For the hydrophilic beads used here, a sessile water droplet of mass $M_d = \rho_l\pi D_0^3 /6$ sinks into the packing within 0.15 seconds. Therefore, the remaining liquid in the residual after the impact would be mixed with grains within this timescale by wetting the opening pores. A complete mixing (Fig.~\ref{f.res_mass}a) results in a mass ratio $m_r = M_r/M_d\approx 11$ for the given volume fraction $\phi$. On the other hand, in the freezing and rapid-freezing regimes, a frozen layer of liquid-grain mixture is formed during spreading or early impact. This frozen layer separates the remaining liquid in the residual and the porous target (Fig.~\ref{f.res_mass}b-c), and few wetting paths are available. In the limiting case of high undercooling degrees, only one or two layers of grains under the spreading droplet contribute to the mixture and, thus, to the residual mass. Consider that the maximum expansion of the droplet indicates the potential area of mixing, the lower bound of $m_r$ is then in the range of \numrange[range-phrase = \text{--}]{1.5}{3} for $D_{m}/D_0=\numrange[range-phrase=\text{--}]{2}{3}$ observed in the experiments. As expected, the measured $m_r$ approaches the upper and lower bounds for small and large undercooling degrees (Fig.~\ref{f.res_mass}d), and it declines significantly in between $ 20\si{K}\lessapprox \Delta T \lessapprox 40\si{K}$. In other words, the impact on hydrophilic granular material appears to become more `hydrophobic' by increasing the undercooling degree of the target. The variation of $m_r$ corresponds to the three undercooling regimes defined by $\Delta\beta$, highlighting the interplay of freezing, liquid-grain mixing and impact dynamics. 

The coupling between solidification and impact dynamics in the studied system is primarily due to the smaller pore dimensions between grains than the droplet size. The solidification process over the pore dimension may be sufficiently fast and comparable to dynamic timescales. In our scenario, surface tension and the inertia of the impacting droplet govern the dynamics. The corresponding timescales, the capillary time $\tau_c=\sqrt{M_d/\gamma}$ and the inertia time $\tau_i$, measure the duration of impact, spreading, retraction, and, of particular interest, the mixing between the droplet and grains. Typically, $\tau_c$ exceeds $\tau_i$. As $\Delta T$ is raised from 0\si{K}, the freezing time across the pore dimension, $\tau_{f}$, gradually decreases and matches first $\tau_c$ and then $\tau_i$. The freezing process of the wetted pores will be further elucidated to clarify the findings presented in Fig.~\ref{f.pheno} and \ref{f.res_mass}.

\subsection{Freezing timescale} 
In a dense packing of spherical particles, the pores can be approximated as capillaries with a diameter of $d_p$. The average distance between grains, $\hat{d}_p=d_p/d_g=\sqrt[3]{1/\phi}-1\approx 0.18$ is used as the dimensionless pore size. For water at $\Delta T\sim 30\si{K}$, solving the conventional Stefan problem yields a sub-millisecond solidification timescale over $d_p$. This, however, is an underestimation. The contact area between grains is negligible compared with their surface area. Therefore, the heat sink under investigation is individual particles of finite sizes. The finite size of grains (the heat sinks) prolongs the freezing time, $\tau_{f}$, and may even prevent the complete solidification of pores. We introduce a dimensionless number $B\equiv{\rho c_p}/(\rho_g c_p^g)$ accounting for the finite size effect, where $c_p$ and $c_p^g$ are the specific heat of ice and grains, respectively. Low-order corrections of $B$ on the freezing process can be made~\cite{Mennig1985}. For brevity, the main results are presented here, with model details available in the supplementary information.

The freezing timescale $\tau_{f}$ accounting for the finite size of grains to the leading order is defined by
\begin{equation}
	\label{eq.tauf}
	\tau_{f} = \frac{\hat{d}_p^2}{2\St} \left(1+\frac 23\frac B\St \hat{d}_p\right)\frac{d_g^2}{\alpha}\quad (\hat{d}_p=d_p/d_g).
\end{equation}
$\alpha$ and $\St \equiv c_p\Delta T/L$ are the thermal diffusivity of the solid phase (ice) and the Stefan number respectively, with the latent heat of phase transition, $L$. The second term within the brackets on the right-hand side of Eq.~\ref{eq.tauf} represents the finite size effect. Its significance is inversely proportional to $\St$ and, thus, to $\Delta T$. For the material under study, this term is approximately 0.5 for $\Delta T\sim 30\si{K}$. As such, its relevance is considered within the examined temperature range. 

The finite size effect also limits the maximum frozen thickness, which is given by $\St/B$ (see SI text for details). Substituting $\hat{d}_p$ with this maximum thickness in Eq.~\ref{eq.tauf} defines an equilibrium time, 
\begin{equation}
	\label{eq.tauF}
	\tau_{F} = ({5\St}/{6B^2})({d_g^2}/{\alpha}),
\end{equation} 
At $\tau_{F}$, grains lose all heat and equilibrate with the solid phase at the melting temperature. 
The complete freezing of the pores thus requires $\tau_{f}\leq\tau_{F}$, or equivalently, $\Delta T\gtrapprox 22.3\si{K}$ for the experimental material, establishing a characteristic undercooling degree $\Delta T_c=22.3\si{K}$. The corresponding freezing time in Eq.~\ref{eq.tauf} for $\Delta T_c$ is $\tau_{f}=4.75\si{ms}$, comparable to the experimental spreading time $\tau_s=1/3\tau_c\approx 5\si{ms}$. 
Hence, solidification in the pore at $\Delta T_c$ completes when the droplet diameter reaches its maximum $D_m$. The capillary pressure on the spread liquid film is the driving force of retraction. Its value, $\gamma D_m^2/D_0^3$, is orders of magnitude lower than the cracking stress of ice $\sim 1\si{MPa}$~\cite{Petrovic2003}. As a result, the whole residual stays at the maximum expansion state (see Fig.~\ref{f.res_mass}b for a sketch). $\Delta T_c$ thus marks the transition to the $\Delta\beta =0$ basin in Fig.~\ref{f.pheno}d, and the theoretically calculated value agrees well with the experimental observation of $20\si{K}$. For $\Delta T\lesssim \Delta T_c$, partial solidification of wetted pores, primarily ice bridges at particle contacts, already diminishes $\Delta \beta$ from its isothermal value. Figure~\ref{f.pheno}d indicates a minimum undercooling degree of $17.5\si{K}$ for retraction suppression, corresponding to a frozen dimension of $0.7d_p$ ($\St/B\approx 0.7$). Nevertheless, the residual mass in Fig.~\ref{f.res_mass} remains unchanged for $\Delta T<\Delta T_c$, as the partially frozen pores permit the residual liquid to sink later.


As $\Delta T$ further increases, the freezing timescale $\tau_{f}$ shortens and reaches the second milestone, the inertia timescale $\tau_i=(D_0+2Z_c)/U$. Here, the substrate deformation, $Z_c$, is taken into account, following Ref.~\citenum{Zhao2015} and \citenum{Zhao2017}. The inertia timescale measures the duration of impact force ~\cite{Soto2014,Eggers2010,Roisman2009} that dominates the initial liquid penetration into the granular substrate~\cite{Zhao2017,Zhao2019}. Another critical undercooling degree $\Delta T_r$ is thus defined by $\tau_{f} = \tau_i$. For $\Delta T>\Delta T_r$, the pores in the substrate are sealed by solidification during early impact, and a layer of icy mixture, with a diameter comparable to $D_0$, forms at the impact center. The droplet then impacts and spreads atop the ice layer. In consequence, the expanding liquid lamella largely dewets from the granular target and retracts later. This sequence of events leads to a proliferation of $\Delta\beta$, reclaiming its isothermal value, as illustrated in Figure~\ref{f.pheno}d. Therefore, $\Delta T_r$ marks the onset of \textit{rapid-freezing} regime.

The computed $\Delta T_r$ lies within the \numrange[range-phrase = \text{--}]{31}{42}\si{K} and increase with $U$, as $\tau_i$ is inversely proportional to $U$. The calculated $\Delta T_r$ matches the observed transition interval between the basin and the plateau of $\Delta \beta$. Nevertheless, experiments do not display the velocity dependence for $\Delta\beta$ against $\Delta T$. One plausible reason is the effective pore dimension sampled during the impact. The pore size in a sphere packing varies with the local packing density or coordination number~\cite{Dollimore1973,Aste2006}. Near the packing surface, the effective packing density is much lower than $\phi$. The effective pore size $\hat{d}_p$ thus varies with the penetration depth of the liquid, $l$, during $\tau_i$ (see Fig.~\ref{f.res_mass}d Inset). For $\hat{d}_p$ to reach the previously used value, the liquid must penetrate at least the first grain layer on the surface, \textit{i.e.,} $l\gtrsim d_g$. It can be shown that $l\sim\sqrt{U}$~\cite{Zhao2017}. Assuming $\hat{d}_p$ is inversely proportional to $l$ to the first order of approximation, then $\tau_{f}\sim\hat{d}_p^2\sim 1/l^2\sim 1/U$. The $U$ dependency in $\tau_{f}$ and $\tau_i$ would cancel out, leading to a much weaker influence of the impact velocity. The justification for this hypothesis will be discussed later.


\subsection{Fragmentation} 

\begin{figure}
	\centering
	\includegraphics{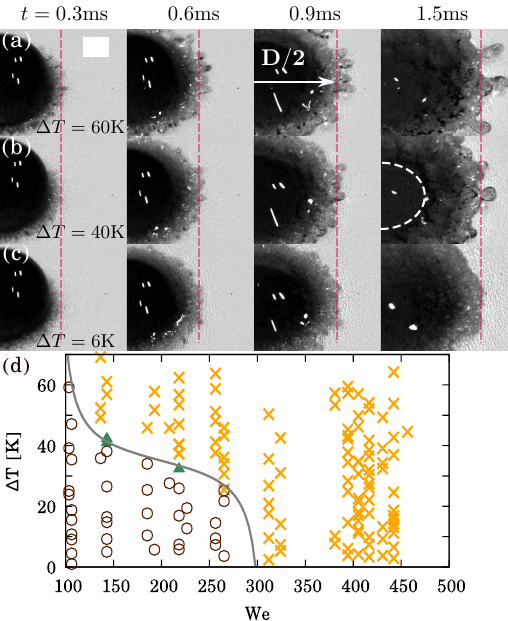}
	\caption{\label{f.early_dyn} The sequential images (background subtracted) depict early-impact dynamics for $U=2.15\si{m/s}$ in the rapid-freezing regime (a-b) and the isothermal regime (c). The dashed curve in (b) indicates the outer edge of the central frozen patch, and $D$ denotes the spreading diameter. The leftmost snapshot in (a) provides a scale bar with dimensions of 1\si{mm} by 1\si{mm}. (d) Fragmentation diagram: $\circ$ denotes no fragmentation, $\blacktriangle$ represents receding break-up, and $\times$ stands for splashing. The solid curve is a visual guide for the fragmentation threshold in Weber number, $\We$.}
\end{figure}

While the retraction degree, $\Delta\beta$, at high undercooling degrees resembles isothermal impacts, the manifested `hydrophobicity', i.e., mixing suppression, in the rapid-freezing regime gives rise to unique dynamic features. A noticeable phenomenon is the enhancement of fingering and fragmentation, which includes both \textit{splashing} during spreading and \textit{receding breakup} during retraction (see Methods and Movie S4 \& S5). Figure~\ref{f.early_dyn}a-c exemplifies this enhancement effect caused by $\Delta T$. In fact, fragmentation is inevitable in the rapid-freezing regime ($\Delta T \geq 40\si{K}$) for $U\geq 1.6\si{m/s}$, equivalent to a Weber number of $\We=\rho_l U^2 D_0/\gamma\gtrsim 140$ (Fig.~\ref{f.early_dyn}d). 

Upon impact, the vertical momentum of the droplet is redirected tangentially along the target surface, causing the ejection of a liquid lamella. Higher ejection velocities promote substantial instability growth of the lamella against surface tension and viscosity. Dimensional analysis suggests that the short-time spreading of the lamella can be described by $D/D_0 \sim (Ut/D_0)^\alpha$, where $D$ denotes the spreading diameter. Previous studies reported $\alpha \approx 0.5$ for impacts on solid surfaces and self-similar theory~\cite{WillemVisser2015,Lolla2022,Mongruel2009,Philippi2016}. In contrast, our experimental data suggests $\alpha\approx 0.3$, regardless of the undercooling degree $\Delta T$ (see Supplementary Information), implying a lower ejection velocity.
We speculate that the slower ejecta velocity is attributed to the reduced impact pressure~\cite{Howland2016}. Under the scenario of impact on a porous target, the maximum pressure during impact is limited by the viscous drag of liquid penetration, potentially an order of magnitude lower than impacts on solid surfaces (see SI for details). This effect is relevant in the initial impact stage (not observable in this study) for the entire $\Delta T$ range, as rapid-freezing must follow the initial wetting of pores, \textit{i.e.},  liquid penetration. Hence, the observed lamella spreading behavior shows no significant variations for different $\Delta T$ values (Fig.~\ref{f.early_dyn}a-c).

The freezing process noticeably influences the later stages. In the rapid-freezing scenarios, the impacting liquid flows atop the frozen layer instead of through the granular target (illustrated in Fig.~\ref{f.res_mass}c). The rim, fed by the pure liquid lamella, is susceptible to capillary instabilities that eventually grow into ligaments. On the other hand, the significant liquid-grain mixing at low $\Delta T$ suppresses the instability due to the increased effective viscosity of the mixture. Receding breakup occurs first as the mixing is diminished by increasing $\Delta T$. While the initial disturbances on the rim could have complex origins, their growth is governed by the local Bond number~\cite{Wang2018}, defined as $\mathrm{Bo}=\rho_l r_m\vert \ddot{D}\vert/\gamma$, where $r_m$ and $\ddot{D}$ denote the rim diameter and rim deceleration. Our experiments indicate that $\mathrm{Bo} \ll 1$ around $t \sim D_0/U$ (see Supplementary Information), thereby highlighting the significance of surface tension in instability development. As a result, the number of ligaments decreases as they merge and grow, as shown in Fig.~\ref{f.early_dyn}a-b.

The relevance of ambient gas in droplet splashing on solid surfaces is well-documented~\cite{Xu2005}. Ambient gas elevates the ejecta and promotes its dewetting from the substrate. However, this mechanism may not be critical in our experiments. The presence of pores in the granular target permits gas filtration under pressure, thus preventing the formation of a continuous gas layer that would otherwise facilitate lift. From the quantitative perspective, we observed a $45\%$ reduction in the splashing threshold, from $\We=257$ to 140, with a modest temperature change of 10\si{K}, as shown in Fig.~\ref{f.early_dyn}d. A recent study on drop impact on cold solid sufraces~\cite{Grivet2023} discovered a similar effect of $\Delta T$ on splashing and proposed that the temperature dependence of gas properties alone cannot explain this level of reduction. 

\subsection{Crater morphology} 
The undercooling degree $\Delta T$ also unexpectedly influences the cratering process. Both the intruder and the substrate deform for droplet impact on granular material. The impact energy, $E_k=M_d U^2/2$, is thus partitioned into two parts, $E_d=(1-\alpha)E_k$ and $E_s=\alpha E_k$, responsible for the deformation of the droplet and the granular layer, respectively. Previous studies have shown that the partition coefficient given by $\alpha=Z_c/(Z_c+D_0/2)$ successfully collapses the characteristic dimensions of the spreading droplet and the crater for various $\phi$, $U$ and liquid properties~\cite{Zhao2015,Zhao2017,deJong2017,deJong2021}. In particular, the maximum crater depth, $Z_c$, follows a scaling law $Z_c/D_0=\Pi\, E_s^{1/3}=\Pi\,\alpha^{1/3}E_k^{1/3}$, where $\Pi$ represents the `softness' of the granular packing~\cite{Zhao2015}. In the case of isothermal impacts, $\Pi$ depends only on $\phi$ for a given grain type. Therefore, with known $E_k$ and $\phi$, $\Pi$ determines $\alpha$ and $Z_c$.

\begin{figure}
	\centering
	\includegraphics[width=8.6cm]{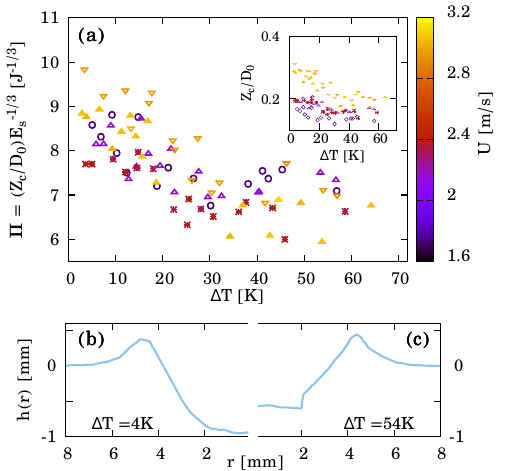}
	\caption{\label{f.zc} (a) The mechanical response of the granular packing to the impact, $\Pi=(Z_c/D_0)E_s^{-1/3}$, follows a master curve depending on $\Delta T$. Color denotes the impact velocity $U$. Inset: the maximum crater depth $Z_c$ dependence on $\Delta T$ for the same dataset. (b-c) are surface height profile $h(r)$ measured at the moment of $h(r=0) = -Z_c$ for two different $\Delta T$, where $r$ is the distance to the impact center. The flat center of $h(r)$ in (c) indicates the frozen patch formed during the early impact. The impact velocity here is 3.06\si{m/s}.}
\end{figure}

For the subfreezing impacts, $\Pi$ facilitates data convergence of $Z_c$ for different impact velocities as well, as illustrated in Fig.~\ref{f.zc}a and its inset. Nevertheless, it is remarkable that $\Pi$ now depends on $\Delta T$, \textit{i.e.,} the packing appears more rigid with increasing $\Delta T$ and saturates for $\Delta T>30\si{K}$. 
The mechanical response of the granular packing is through extensive force chains~\cite{Cates1998}. In the rapid-freezing regime, only the first layer of grains on the surface within an area of $\pi D_0^2/4$ underneath the impacting droplet is frozen into a patch (Fig.~\ref{f.early_dyn} and Fig.~\ref{f.zc}c). The solidification process thus does not change the area for vertical momentum transfer. Instead, it alters the robustness of force chains. 
We propose that the rigidity enhancement stems from shielding the shear stresses of droplet impact. 
Studies have revealed that a localized shear band agitates the breaking of force chains outside the sheared zone in a granular packing~\cite{Nichol2010,Reddy2011}. In other words, shearing on the surface softens the bulk. This resistance-reduction strategy has been employed by some species of flowering plants, such as \textit{Philodendron appendiculatum}, for their seeds to burrow into the earth effectively~\cite{Jung2017}. Droplet impact induces intense shearing on the substrate~\cite{Philippi2016,Sun2022}, a known cause of soil erosion~\cite{Kinnell2005,NouhouBako2016}. The frozen layer formed during the early impact separates the impacting liquid and spreading lamella from the packing surface. It thus shields the packing surface from the shear stresses during impact. As a result, the granular target appears more rigid, and $\Pi$ and $Z_c$ decrease.

\section{Discussion} 
\begin{figure}
	\centering
	\includegraphics{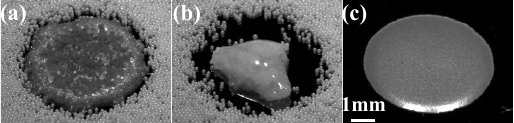}
	\caption{\label{f.hex} The final solidification morphology of hexadecane droplet impact on a monolayer of beads on a silicone wafer at $\Delta T=3\si{K}$ (a) and 14\si{K} (b). 
		(c) Morphology of a hexadecane droplet impact on a silicone wafer at $\Delta T = 21\si{K}$.
	}
\end{figure}

We have explored the influence of solidification on the droplet dynamics during the impact on cold granular material. In the rapid-freezing regime, the formation of a frozen layer at early-impact inhibits liquid penetration and shields the target from shear stresses. As a result, the hydrophilic impact appears hydrophobic; the bulk solidification is postponed; the target deforms less. Since this rapid solidification process primarily involves surface grains, a monolayer is sufficient for the same effects, as confirmed by preliminary experiments (see Fig.~\ref{f.hex}). Our research thus suggests a potential strategy for protection against icing and impact stresses on cold substrates. Current anti-icing surfaces often use superhydrophobic micro-textures to reduce droplet contact time~\cite{Schutzius2015,Mishchenko2010,Golovin2016,Kreder2016}. However, the possibility of droplet penetration into these textures, leading to failure, is a concern~\cite{Maitra2014}. This risk may be mitigated by adjusting the pillar-pitch dimension ratio, similar to the roles of $d_p$ and $d_g$ in the rapid freezing process studied. The monolayer configuration could improve comprehension of the rapid-freezing process and its practical aspects. For instance, it allows for the measurement of grain motion and the depth of liquid infiltration during impact, which helps to determine whether the lack of dependence on velocity ($U$) for the onset temperature of the rapid-freezing process is specific to impacts on subfreezing packings.

Lastly, the effect of grain size, $d_g$, needs further attention. It is worth noting that the two critical freezing temperatures, $\Delta T_c$ and $\Delta T_r$, are defined distinctly. $\Delta T_c$ denotes the onset of complete solidification, a prerequisite for rapid-freezing phenomena. Therefore, $\Delta T_r<\Delta T_c$ is unrealistic. The value of $\Delta T_c$, given by $\tau_{f}=\tau_{F}$, is independent of $d_g$, since both $\tau_{f}$ and $\tau_{F}$ scale with $d_g^2$ (Eq.~\ref{eq.tauf} and \ref{eq.tauF}). However, $\Delta T_r$, the onset of the rapid-freezing regime, increases with $d_g^2$. Therefore, a critical grain size, $d_g^*$, exists where the rapid freezing process occurs at $\Delta T_r=\Delta T_c$. For the material investigated, $d_g^*$ is approximately 80\si{\micro\meter}. Reducing the grain size further does not lower the undercooling threshold for rapid-freezing and the shielding effect. However, smaller grains may be beneficial to prevent liquid penetration during impact. It is noteworthy that the inertia of individual grains decreases significantly at such small sizes, and the timescales of their motion driven by the impact stresses may become relevant, which causes rearrangements of pores. Simulations, theories and experiments give a shear stress scale, $\sigma\sim\sqrt{\eta\rho_l U^3/D_0}$, for droplet impact on solid surfaces~\cite{Roisman2009,Eggers2010,NouhouBako2016,Philippi2016,Sun2022}, with $\eta$ as the dynamic viscosity of the impacting liquid. The sliding timescale of grains under shear, $\tau_g=\sqrt{2\rho_g d_g^2/\sigma}$, can thus be estimated. If $\tau_g<\tau_f$, the scenario of freezing stationary pores becomes invalid. Hence, when applying the freezing model to small or light grains, $\tau_f$ should be compared with the smaller one of $\tau_g$ and $\tau_i$ to determine the onset undercooling degree of the rapid-freezing process. 

\bibliography{impact_cold_grain}

\pagebreak

\appendix
\section{Freezing with finite-size grains} 
The freezing front propagation is usually modeled as a Stefan problem for droplets impacting on an undercooled solid substrate. We consider the solidification of the impacting liquid within the pores of a granular target. The contact area between individual grains is negligible compared to their surface. Therefore, the finite size of individual grains as a heat sink needs to be considered. 

\begin{figure}
	\centering
	\includegraphics{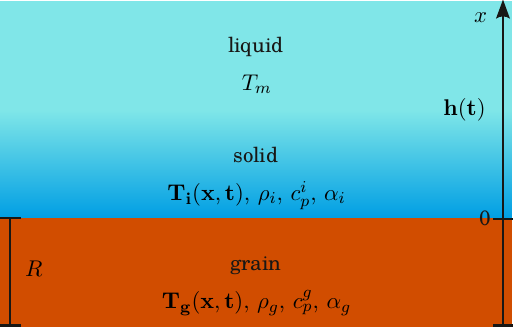}
	\caption{\label{f.model}A schematic of the freezing model.}
\end{figure}

We first establish a one-dimensional freezing model. The liquid is at the melting temperature, $T_m$. $h(t)$ denotes the position of the liquid-solid interface. It matches the grain surface at $x=0$ initially. Thermal diffusion constantly transmits the latent heat associated with the phase transition, $L$, to the heat sink (the grain). There is no heat flux across the bottom of the heat sink at $x=-R$. We consider here that the freezing is symmetric between neighboring particles. $R$ is then the particle radius, $d_g/2$, and the \textit{complete solidification} is marked by the moment of $h(\tau_{f})=\hat{d}_p R$. Here, $\hat{d}_p$ denotes the dimension ratio between the pore and the grain, the same as in the main text. The problem is to solve the temperature in the solid phase, $T_i(x,t)$, that in the grain, $T_g(x,t)$, and the interface position, $h(t)$. Please refer to Figure~\ref{f.model} for the model configuration.

The problem now has a characteristic length scale $R$. The initial temperature of grains, $T_{g0}$, is taken as the reference temperature. 
Scales of temperature and time are chosen as $T_m-T_{g0}$ and $R^2/\alpha_i$, where $\alpha_i = {k_i / (\rho_i c_p^i)}$ is thermal diffusivity of the solid phase (ice). The subscript $g$ is used for the corresponding thermal constants of the heat sink. Henceforth, quantities are non-dimensionalized with the scales chosen above. The same symbols are used unless defined otherwise.

The dimensionless heat equations and boundary equations are  
\begin{subequations}
	\label{eq.stefan}
	\begin{align}
		\pdiff{{T_i}}{t} & = \pdiff[2]{{T_i}}{x} \\
		{T_i}({h}(t),t) & = 1\\
		\pdiff{T_i}{x}\Big\vert_{x=h(t)} &= \frac 1\St \dot{h}\\
		\pdiff{T_g}{t} &= \frac BC \pdiff[2]{T_g}{x}		\\
		\pdiff{T_g}{x}\Big\vert_{x=0} &= C \pdiff{T_i}{x}\Big\vert_{x=0}\\
		\pdiff{T_g}{x}\Big\vert_{x=-1} &= 0
	\end{align}
\end{subequations}
with $\St = c_p^i(T_m-T_{g0})/L$, $B=\displaystyle \frac{\rho_i c_p^i}{\rho_g c_p^g}$ and $C = k_i/k_g$. Equation~\ref{eq.stefan}a-f are to be solved with the initial conditions: 
\[
T_i(x,0) = 1;\quad T_g(x,0) = 0;\quad h(0) = 0;\quad \pdiff{T_i(x,t=0)}{x} = 0; \quad \pdiff{T_g(x,t=0)}{x} = 0;
\]
Compared to the conventional Stefan problem, Equation~\ref{eq.stefan}d-f are the additional thermal equations and boundary conditions for the heat sink (grains). In particular, Eq.~\ref{eq.stefan}f sets it apart from the infinite heat sink problem~\cite{Thievenaz2019}, where a self-similar solution can be constructed. A criterion can be established based on the self-similar solution (for instance, in Ref.~\citenum{Thievenaz2019}). For $\St\ll 1$, the diffusion coefficient for freezing front development is proportional to $\St^2\,\alpha_i$. If the ratio $\St^2\,\alpha_i/\alpha_g$ is smaller than $(d_p/d_g)^2$, the boundary condition Eq.~\ref{eq.stefan}f, and thus the lengthscale $d_g$, becomes relevant.

If the heat transfer within the grains is significantly faster than that in the ice, \textit{i.e.,} $C\sim 0$, Eq.~\ref{eq.stefan}d-f are reduced to 
\begin{equation}
	\label{eq.reduced_boundary}
	F(t) = \frac 1B \pdiff{T_i}{t}\Big\vert_{x=0}.
\end{equation}
We define $F(t) \equiv\pdiff{T_i}{x}\Big\vert_{x=0}$ for convenience. $0\leq F(t)\leq 1$ increases with time. 
In the following, we obtain a low-order approximation solution for this reduced problem and then justify its usage to estimate the typical freezing time in the main text.

When solving Eqs.~\ref{eq.stefan}a-c and Eq.~\ref{eq.reduced_boundary}, we adapt an approximation proposed by Mennig~\cite{Mennig1985}, where gradients of $T_i$ and $\pdiff{T_i}{x}$ are assumed linear. It gives
\begin{subequations}
	\label{eq.reduced_approx}
	\begin{align}
		2h-2\frac \St B F + (1-F) \St h &= 0\\
		\frac{1}{4\St }\dot{h^2} + \frac{1}{2B}h\dot{F} + F &= 1
	\end{align}
\end{subequations}
In the situation of $\St \ll 1$, the last term in Eq.~\ref{eq.reduced_approx}a is neglected and $\displaystyle F \approx \frac{\St +2}{2} \frac B\St  h \approx \frac B\St  h$. This approximation is consistent with the lowest order correction of $F$ for finite $B$ in the perturbation method, and the corresponding solution of the solidification front is then 
\[h(t) + \frac{\St }{B}\ln\left(1-\frac B\St  h(t)\right) = -Bt.\]
This solution can be further expanded into polynomials of $B$. The first-order correction reads
\begin{equation}
	\label{eq.reduced_sol}
	h^2+\frac 23\frac{B}{\St } h^3 = 2\St\, t.
\end{equation}

The $h(t)$ solution is only valid for the duration corresponding to $0\leq F\leq 1$. For the parameter range studied here, $\St \approx 0.2$ (for $\Delta T=T_m-T_{g0}\sim 30\si{K}$) and $B \approx 0.7$ (see Tab.~\ref{t.properties}) and $h=\St /B\approx 0.3$ when $F=1$. Therefore, the cubic term of $h$ in Eq.~\ref{eq.reduced_sol} is comparable to the quadratic term, indicating the significance of the finite size effect. One may define a freezing timescale $\hat{\tau}_{f}$ corresponding to $h=\hat{d}_p$ via Eq.~\ref{eq.reduced_sol}, where $\hat{d}_p$ is the dimensionless length scale of pores in the granular substrate. In addition, a timescale for the validation of the solution, $\tau_{F}$, can be defined corresponding to $F=1$:
\begin{equation}
	\tau_{F} = \frac{5\St }{6B^2},
\end{equation}
so that complete freezing is unrealistic when $\tau_{f}>\tau_{F}$.

A similar procedure can be applied to Eqs.~\ref{eq.stefan}a-f, considering the finite heat conductivity of grains (finite $C$). The approximate solution is 
\begin{equation}
	\label{eq.full_sol}
	\frac{P}{Q}\tilde{h}(t) + \left(\frac{C}{4}+\frac{P}{Q^2}\right)\ln\Big(1-Q \tilde{h}(t)\Big) = -Bt,
\end{equation}
where $P=B(1/\St +1/4)$ and $Q = B(1/\St +1/2)$. $\tilde{h}(t)$ in Eq.~\ref{eq.full_sol} denotes the interface position.

\begin{figure}
	\centering
	\includegraphics[width=8cm]{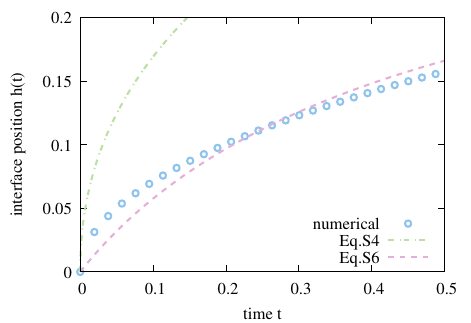}
	\caption{\label{f.num_stefan} Comparison of the low-order approximation and numerical solution of the freezing on a finite heat sink problem. The values of dimensionless parameters are $\St =0.2$, $B=0.7$ and $C=1$.}
\end{figure}

For the material used in experiments, we have $C\approx 1$ (see Tab.~\ref{t.properties}).
Equations~\ref{eq.reduced_sol} and~\ref{eq.full_sol} are compared with the numerical solution of Eqs.~\ref{eq.stefan}a-f in Figure.~\ref{f.num_stefan} with dimensionless numbers selected. The solution in Eq.~\ref{eq.full_sol} captures the freezing process semi-quantitatively, slightly underestimating initially and overestimating for a longer time. When obtaining Eq.~\ref{eq.reduced_sol}, the thermal conductivity of the heat sink is ignored ($C=0$). Therefore, it underestimates the freezing time $\tilde{\tau}_f$ by a factor of about 4. Namely, for the typical value of $\St $, $B$ and $C$, the freezing time can be calculated by Eq.~\ref{eq.reduced_sol} with a factor $\mathcal{D}$:
\[ \frac{\tilde{\tau}_{f}}{\mathcal{D}} = \hat{\tau}_f = \frac{\hat{d}_p^2}{2\St } \left(1+\frac 23\frac B\St \hat{d}_p\right) \text{with } \mathcal{D}\approx 4
\]
To calculate the \textit{dimensional} freezing time $\tau_{f}$ using Eq.~\ref{eq.reduced_sol}, one can simply replace the characteristic timescale, $R^2\alpha_i$, with $\mathcal{D} R^2/\alpha=d_g^2/\alpha$. This results in Equation~1 from the main text. This substitution is used for its conciseness and brevity but does not affect the critical undercooling degree $\Delta T_c$, which remains unchanged regardless of the characteristic timescale chosen. The value of $\Delta T_c$ is determined only by the finite size effect (a finite $B$), not by the finite thermal conductivity.

The parameters used for the calculation are given in Tab.~\ref{t.properties}
\begin{table*}[h]
	\centering
	\caption{\cite{haynes2016crc,biddle2013thermal,Youngblood1988,Degueldre2003}}
	\label{t.properties}
	\renewcommand\arraystretch{}
	\begin{tabular}{*{8}{c}}
		\toprule
		\textbf{Material} & \textbf{T \SI{}{\degreeCelsius}} & \textbf{$\rho$ \SI{}{\ensuremath{kg/m^{3}}}} & \textbf{$c_{p}$
			\SI{}{kJ/(kg \cdot K)}} & \textbf{$k$ \SI{}{W/(m \cdot K)}} & \textbf{L \SI{}{kJ/kg}} \\
		Droplet& 0.1 & 999.8 & 4.22 & 0.561 & 333.4 \\
		Ice & -0.1 & 916.7 & 2.1 & 2.16 & 333.4 \\
		\ch{ZrO2} beads & 0 & 6050 & 0.48 & 2.3\footnote{of single crystal in Ref.\citenum{Degueldre2003}} & \\
		\botrule
	\end{tabular}
\end{table*}

\section{Early spreading behavior}
\begin{figure}
	\centering
	\includegraphics{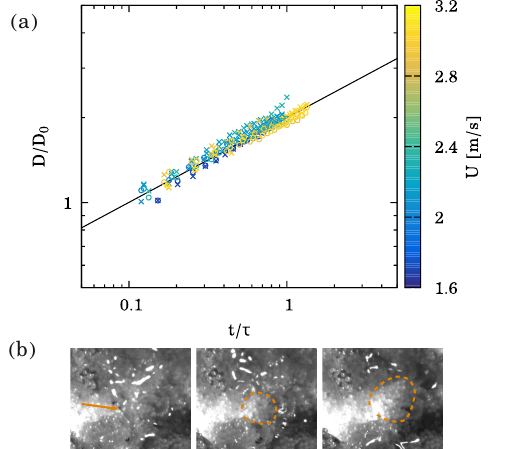}
	\caption{\label{f.early_spr} (a) The spreading diameter, $D$, in the isothermal ($\Delta T \approx 3\si{K}$) and the rapid-freezing regime ($\Delta T \approx 50\si{K}$) are denoted by $\circ$  and $\times$, respectively. The solid line is the power law, ${D/D_0}= 2(Ut/D_0)^{0.3}$. 
		(b) Sequential snapshots illustrate the rupture of a portion of the pure liquid film after spreading. The arrow in the leftmost panel indicates the point of rupture. The dashed curve denotes the edges of the liquid film after 0.7 ms and 1.6 ms. This phenomenon often happens in the transition region from freezing to rapid-freezing, where the rim is partially mixed with grains and frozen. The unmixed portion undergoes ruptures, while the frozen part is anchored on the target surface.
	}
\end{figure}

We measured the spreading diameter of the droplet, $D$, at the early stage of the impact. When using $D_0$ and $\tau = D_0/U$ as the characteristic length and time scale for non-dimensionalization, data for all $U$ collapses on a master curve [Figure~\ref{f.early_spr}a], $\hat{D}= a\hat{t}^{b}$. Our data suggests $a=2.028 \pm 0.008$ and $b = 0.290 \pm 0.006$. Note that the conventional inertia time scale is used here to compare with other studies, as for $t<\tau$, the granular target does not deform much. Using $\tau_i=(D_0+2Z_c)/U$ instead does not significantly alter the scaling relation. This observation does not disprove the well-known $t^{1/2}$ law for the initial wetting area on the substrate, which typically occurs for $t/\tau < 0.1$.

Compared to $\alpha\sim 0.5$ for droplet impact on solid with similar $U$, the dependence of $\hat{D}$ on time here is weaker, indicating a lower ejection velocity. The slower ejecta may be due to the reduced maximum pressure during impact, which is bounded by the viscous drag of liquid penetration here $\sim \mu U/d_p \sim (D_0/d_p) (\rho_l U^2/\Rey)$, with $\mu$ as the dynamic viscosity. For $\Rey > 10^2$, this pressure $\lesssim \rho_l U^2$ becomes considerably lower than that for impact on solid surfaces $\sim 10\rho_l U^2$.

From the fitted relation $\hat{D}= a\hat{t}^{b}$, we calculate the deceleration of the rim (edge of the expanding lamella), $\ddot{R}=\ddot{D}/2$. Here, $\ddot{D}$ represents the second time derivate of $D$. We consider the moment of $\hat{t} \sim \mathcal{O}(1)$. Then, $\vert\ddot{R}\vert\approx 0.2 {U^2}/{D_0}$ is between 170 and 530 \si{m/s^2} for the range of $U$ explored. The rim diameter $r_m$ is measured 150\si{\micro\meter} at $\hat{t}=1$ in the experimental videos. This length scale of $r_m$ can be additionally confirmed. In the rapid-freezing regime, the extended lamella largely dewets from the granular target and ruptures during receding under certain conditions, as shown in Figure~\ref{f.early_spr}b. Using the Taylor-Cuilick relation and the rupture velocity, we estimate the lamella thickness of 100\si{\micro\meter}. At the early spreading stage, $r_m$ is of the same order of magnitude as the lamella. All together the Bond number $\mathrm{Bo}=\rho_l r^2_m\vert \ddot{R}\vert/\sigma$ is of the order of 0.05 for lowest $U$ and 0.16 for the highest $U$.


.
%

\end{document}